# Time-Domain Doppler Biomotion Detections Immune to Unavoidable DC Offsets

Qinyi Lv, Lingtong Min, Congqi Cao, Shigang Zhou, Deyun Zhou, Chengkai Zhu, Yun Li, Zhongbo Zhu, Xiaojun Li and Lixin Ran

*Abstract*—In the past decades, continuous Doppler radar sensor-based bio-signal detections have attracted many research interests. A typical example is the Doppler heartbeat detection. While significant progresses have been achieved, reliable, time-domain accurate demodulation of bio-signals in the presence of unavoidable DC offsets remains a technical challenge. Aiming to overcome this difficulty, we propose in this paper a novel demodulation algorithm that does not need to trace and eliminate dynamic DC offsets based on approximating segmented arcs in a quadrature constellation of sampling data to directional chords. Assisted by the principal component analysis, such chords and their directions can be deterministically determined. Simulations and experimental validations showed fully recovery of micron-level pendulum movements and strongly noised human heartbeats, verifying the effectiveness and accuracy of the proposed approach.

## I. Introduction

CONTINUOUS-WAVE Doppler radar sensors (CDRS) are considered to have a wide range of potential applications. Typical examples include mechanical vibration measurement [1], victims search and rescue [2], through-wall life detection [3] and fall detection [4]. Due to the unique properties such as the simple structure, high sensitivity and ultra-narrow spectrum occupancy, CDRSs are especially suitable for non-contact bio-signal detections. So far, there have been many progresses achieved in architectures and algorithms, to linearly demodulate the Doppler phases from signals scattered by moving bio objects, especially in time domain [5]-[25].

Initially, the small angle approximation was used to demodulate heartbeat signals based on zero-IF CDRSs [8]-[9]. However, it suffers from null-point issues depends on the detecting distance. To overcome this problem, the direct arctangent-based phase demodulation was used [10]. This approach is able to obtain more accurate and stable measurements compared to the small-angle approximation. However, the arctangent function mathematically has a native codomain range of ($-\pi/2$, $\pi/2$), which will cause discontinuities in demodulated signals. It also suffers from the flicker noises and direct current (DC) offsets unavoidable in zero-IF receivers [11].

To solve the above problems, improved phase demodulation algorithms such as the differentiate and cross-multiply (DACM) and the arcsine algorithms were proposed for digital-IF, quadrature-structured CDRSs [12]-[13]. The use of the digital-IF structure completely removes the flicker noises, the circuit DC offsets and quadrature imbalances, and the application of the DACM and the arcsine algorithms eliminates the phase ambiguity caused by the direct arctangent algorithm. Combing algorithms such as the gradient descent [14] and the Levenberg-Marquardt center estimation [15] algorithms that can determine and dynamically trace the DC offsets caused by unavoidable stationary scatterers, Doppler phases induced by both small- and large-scale motions can be linearly demodulated. So far, such algorithms have been successfully used in various applications. Examples include human heartbeat detections [16]-[17], Doppler cardiograms [18], hand gesture recognitions [19] and indoor localizations [20].

Although DACM and arcsine algorithms have demonstrated quite good demodulation performance, they still face challenges in the robust detection of weak biomotions such as human heartbeats. For such motions, a small error of the determined DC offset could result in a considerable distortion of the demodulated bio-signal. When the signal-to-noise ratio (SNR) of a signal detected by a CDRS is not satisfactory, the demodulated motion can be easily overwhelmed by noises. Finally, both the determination and the dynamic tracing of DC offsets would cost considerable computational time, unsuitable for real-time detections.

In this paper, we propose a novel demodulation algorithm that is immune to DC offsets for detecting weak biomotions such as human heartbeats. It approximates segmented arcs in the quadrature constellation of sampled data to directional chords. Assisted by the principal component analysis (PCA), such chords and their directions can be deterministically determined. Simulations and experimental validations showed the fully recovery of micron-level pendulum movements and strongly noised human heartbeats. By doing so, not only the

Manuscript received January 1, 2021. This work is supported in part by the National Natural Science Foundation of China under Grants 61906155, 61771395, 61771421, and 62071417. (*Corresponding authors: Qinyi Lv and Lixin Ran.*)

Q. Lv, L. Min, C. Cao, S. Zhou and D. Zhou are with the School of Electronics and Information, Northwestern Polytechnical University, Xi'an 710072, China (e-mail: lvqinyi@ nwpu.edu.cn).

C. Zhu and L. Ran are with the Laboratory of Applied Research on Electromagnetics (ARE), Zhejiang University, Hangzhou 310027, China (e-mail: ranlx@zju.edu.cn).

Y. Li, Z. Zhu and X. Li are with the National Key Laboratory of Science and Technology on Space Microwave, Xi'an 710100, China.

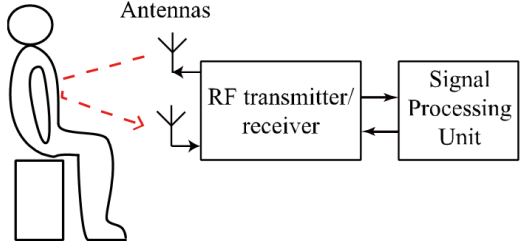

Fig. 1. Diagram of the bio-signal detection based on a digital-IF CDRS.

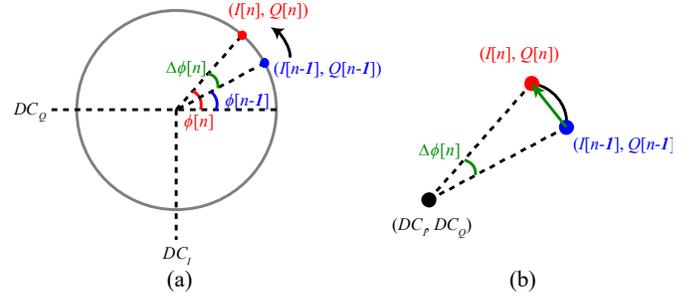

Fig. 2. Mathematical model for calculating phase differences. (a) Conventional arc-based method. (b) Proposed chord-approximation method.

time-consuming DC offset determination and tracing is completely avoided, the demodulation can be robustly carried out even if the noise is strong enough to disable the aforementioned phase-demodulation algorithms, exhibiting promising potentials in applications of biomotion detections.

## II. THEORY

### A. Problem description

The block diagram of a CDRS used to detect small-scale bio-signals is shown in Fig. 1. As a continuous-wave radar sensor, its transmitting and receiving channels should be coherent. The digital-IF CDRS transmits a CW to illuminate the target. Scattered by the biomotion with a time-varying displacement $x(t)$, the received signal is quadrature demodulated in digital domain. The digitized in-phase ($I$) and quadrature-phase ($Q$) outputs can be described as

$$I[n] = A_I \cos\left[\theta_0 + \frac{4\pi x[n]}{\lambda} + \varphi_0[n]\right] + DC_I[n], \quad (1)$$

$$Q[n] = A_Q \sin\left[\theta_0 + \frac{4\pi x[n]}{\lambda} + \varphi_0[n]\right] + DC_Q[n]. \quad (2)$$

where $n$ represents the $n$th sampling in digital domain, $\varphi_0[n]$ denotes the residual phase noise of the receiver, $\theta_0$ represents the phase shift due to the initial distance between the subject and CDRS, $A_I$, $A_Q$ denote the amplitudes of the detected $I/Q$ signals, and $DC_I[n]$, $DC_Q[n]$ are the DC offsets due to environmental stationary scattering and the potential crosstalk between transmitting and receiving channels, respectively. It should be noted that apart from the above DC offsets, Doppler signals due to biomotions also contain useful DC components [6]. Therefore, the DC components in (1) and (2) cannot be simply filtered [26]. Otherwise, the biomotions would be recovered with inevitable distortions.

Based on (1) and (2), the motion $x[n]$ is proportional to the phase $\phi[n]$, which can be mathematically solved by

$$\phi[n] = \arctan\frac{\left[Q[n]-DC_Q[n]\right]/A_Q}{\left[I[n]-DC_I[n]\right]/A_I} - \theta_0 - \varphi_0[n] \quad (3)$$

where the residual phase noise $\varphi_0[n]$ can be neglected when the transmitting and receiving signals are coherent [21]. Since only relative motions are concerned, the initial phase $\theta_0$ can also be ignored. In digital domain, an ideally symmetrical quadrature down-conversion can be obtained, and thus the imbalance between $A_I$ and $A_Q$ can be eliminated. As a result, the phase demodulated can be simplified as

$$\phi[n] = \arctan\frac{Q[n]-DC_Q[n]}{I[n]-DC_I[n]} \quad (4)$$

It is can be seen that even the $I/Q$ signals can be accurately obtained, the linearity and accuracy of the phase demodulation can be directly affected by the DC offsets. It means that in order to obtain satisfactory results based on a conventional phase demodulation algorithm, the DC offsets $DC_I[n]$ and $DC_Q[n]$ have to be accurately determined.

Fig. 2(a) illustrates the constellation diagram of two adjacent $n$th and ($n$-1)th quadrature data sampled by a CDRS. Detected within a short period, the corresponding two points can be considered to locate on a circle whose center is ($DC_I$, $DC_Q$). In practice, this circle center could be dynamically moved due to large-scale motions or changes of stationary environmental scatterings. In this case, the detected quadrature signal has to be segmented into short segments, to dynamically determine ($DC_I$, $DC_Q$) based on optimizations such as the gradient descent used in [14] and the random sample consensus (RSC) used in [18]. Then, demodulation algorithms such as the DACM [12] can be used to retrieve the motion $x[n]$.

However, while such optimizations can be very time consuming, the accuracy of the determined circle centers severely infected by noises. According to (4), this will introduce systematic errors in the demodulated biomotions.

### B. Chord Approximation

To completely solve this problem, we propose an approach that abandons to determine circle centers. First, we increase the sampling rate to reduce the length of the arc between adjacent sampling points, so that the chord shown in Fig. 2(b) approximates this arc. Note that increasing sampling rate is not a burden for detecting biomotions, whose period is normally second-level.

Under this approximation, the length of the chord will be proportional to $\Delta\phi[n]$. It means that if the physical amplitude of the demodulated motion is not concerned, a normalized motion can always be obtained by simply treating the length of the chord as $\Delta\phi[n]$. By doing so, while all the shortages of the dynamic DC-offset determination and tracing can be completely avoided, concise analytical equations can be derived to calculate $\Delta\phi[n]$. Without any necessity of iterative optimization, it saves considerable computation time, suitable for real-time biomotion detections.

Accordingly, the equations of the proposed chord approximation can be derived as

$$\Delta\phi[n] \approx sign[n]\frac{\sqrt{(Q[n]-Q[n-1])^2+(I[n]-I[n-1])^2}}{R} \quad (5)$$

$$\begin{aligned}x[n] &= \frac{\lambda}{4\pi}\phi[n] \\ &\approx \frac{\lambda}{4\pi R}\sum_{k=2}^{n}\left(sign[k]\sqrt{(Q[k]-Q[k-1])^2+(I[k]-I[k-1])^2}\right)\end{aligned} \quad (6)$$

respectively, where $R$ corresponds to amplitudes $A_I$ or $A_Q$, used in normalizing the demodulated motion, $sign[n]$ represents a sign bit with 1-bit values 1 or -1, used to determine cumulative addition or subtraction of the phase $\phi[n]$, respectively.

By approximating the arcs as chords, as seen in (6), the biomotion $x[n]$ has nothing to do with the DC offset. Therefore, the determination and tracing of circle centers is no longer needed, and the motion $x(t)$ normalized by $\lambda/4\pi R$ can be directly obtained. In practice, $x(t)$ can also be normalized by the peak amplitude of the detected signal.

*C. Sign bit determination*

For the proposed chord approximation, a robust determination of $sign[n]$ would play a key role. In this work, it was based on a "directional line" fitted by a set of sampling points. Its direction is defined from the start point to the end point of the chord approximated for each segmented quadrature data. If the angle included between two adjacent sampling points and this directional line is smaller than 90 degrees, i.e., the inner product is positive, this sign bit will be specified as +1. Otherwise, it will be specified as -1.

Therefore, the fitting of the directional line is crucial to the chord-approximation method. Mathematically, both the least square (LS) and the PCA methods can be used to fit the directional line. In this work, we chose the PCA so that it can robustly fit the directional lines for continuously sampled data located in any area of the quadrature constellation.

Figs. 3(a) and 3(b) illustrate the fitting of four sampling points with the LS and the PCA methods, respectively. The LS method minimizes the vertical distances between discrete points and the directional line, and thus it could be failure for points scattered close to the vertical direction. In comparison, the PCA method minimizes the perpendicular distance to the directional line, and thus it would work for arbitrary data.

Mathematically, PCA can be applied as follows. First, for a set of sampling data $(I[1], Q[1]), (I[2], Q[2]),\ldots (I[N], Q[N])$, the mean of the $I/Q$ signals can be subtracted respectively, i.e.,

$$I_0[i] = I[i] - \frac{1}{N}\sum_{i=1}^{N}I[i]$$

$$Q_0[i] = Q[i] - \frac{1}{N}\sum_{i=1}^{N}Q[i], i=1,2,\cdots,N \quad (7)$$

Then, a signal matrix $\mathbf{M}$ can be composed as

$$\mathbf{M} = \begin{bmatrix} I_0[1] & I_0[2] & \cdots & I_0[N] \\ Q_0[1] & Q_0[2] & \cdots & Q_0[N] \end{bmatrix} \quad (8)$$

Next, singular value decomposition (SVD) decomposition can be employed to matrix $\mathbf{M}*\mathbf{M}^T$, where $\mathbf{M}^T$ is the transpose of $\mathbf{M}$, and thus

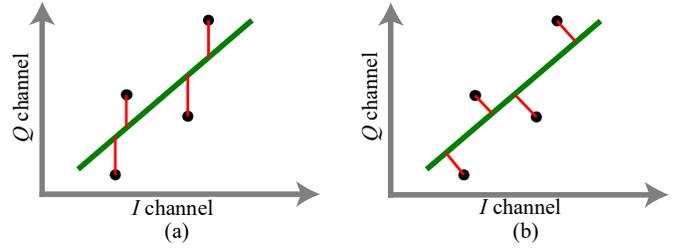

Fig. 3. Fitting of the directional line based on (a) the LS and (b) the PCA methods.

$$\mathbf{M}*\mathbf{M}^T = \mathbf{U}*\mathbf{S}*\mathbf{V}^T \quad (9)$$

where $\mathbf{U}$, $\mathbf{S}$ and $\mathbf{V}$ represent the left-singular vectors, the singular value matrix and the right-singular vectors, respectively.

According to the PCA theory, the normal vector $\mathbf{n}$ of the fitting line points to the characteristic direction corresponding to the smallest eigenvalue of the dataset matrix $\mathbf{M}*\mathbf{M}^T$, which is symbolled as $\mathbf{n} = [n_1, n_2]$.

Finally, the directional line can be determined by

$$y = k*x + b \quad (10)$$

where

$$k = -n_1/n_2 \quad (11)$$

and

$$b = \frac{1}{N}\sum_{i=1}^{N}Q[i] - k*(\frac{1}{N}\sum_{i=1}^{N}I[i]) \quad (12)$$

Compared with the time-consuming, iterative optimizations used in determining and tracing the dynamic DC offsets, the major computational cost of the PCA is the SVD for $\mathbf{M}$. Assisted by the PCA, real-time detections of biomotions become realistic.

III. SIMULATION

According to the above discussions, the key of the proposed approach is to correctly determine the sign bit. To compare the performance between the LS and the PCA methods in determining correct sign bits, four segments of 1.3-Hz, strongly noised sinusoidal signals were generated based on (1) and (2), to simulate heartbeat signals. In the generation, the heartbeat amplitude, frequency and the sampling rate were set as 1 mm, 1.3 Hz and 90 Hz, respectively, $A_I$ and $A_Q$ were normalized as unity, respectively, $\lambda$ was set as 12.5 mm, corresponding to the frequency of 24 GHz, and the initial phase $\theta_0$ was set as 0º, 30º, 60º and 90º for different segments, respectively.

First, the red sampling points comprising the $I/Q$ constellations in panels I, II, III and IV in Fig. 4 were added with Gaussian noises, so that the signal-to-noise ratios (SNRs) of such signals are all 3 dB. Before the noises were added, all the original sampling points locate on four ideal arcs, as shown in gray in each panel. In all the panels, the angles included by the chords of such ideal arcs and the $x$ direction denoted in each panel are 0º, 30º, 60º and 90º, respectively. Here the $x$ direction is parallel to the $I$-channel axis.

As seen in Fig. 4, with the 3-dB SNR, the actual DC offsets, i.e., the circle centers denoted as black dots, have located among the raw sampling points. In this case, any effort trying to locate the circle centers will fail. As examples, the brown dots in Fig. 4 show the circle centers optimized at (-0.02, 0.59), (0.60, 0.90), (0.92, 0.62) and (0.62, -0.02) based on the gradient descent algorithm proposed in [14], respectively. They are obviously incorrect. Instead, by approximating the arcs as chords, as seen in (6), the biomotion $x[n]$ has nothing to do with the DC offset, meaning that it is no longer needed to localize the DC offset.

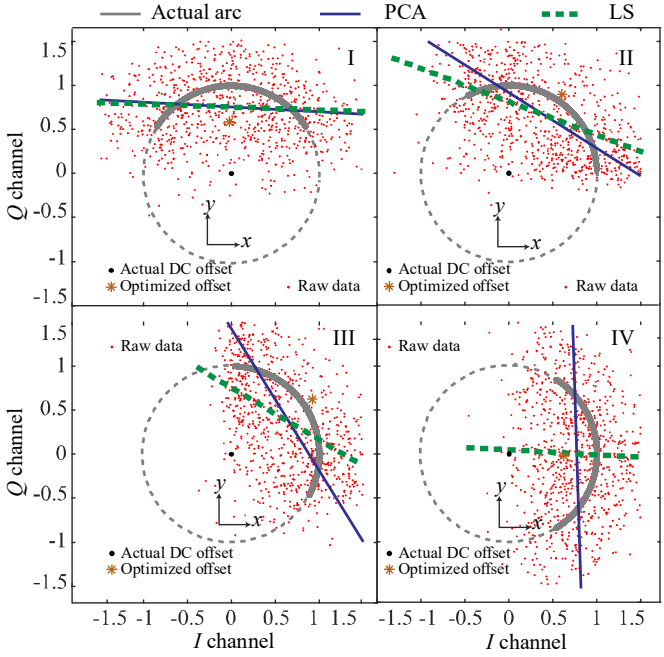

Fig. 4. Simulation comparison between the LS and the PCA methods.

For the noisy data shown in each panel, the PCA and LS methods were separately used to fit the corresponding directional lines. The obtained lines are shown in each panel in solid blue and dashed green, respectively.

In Panel I, it is seen that the directional lines fitted by the PCA and LS are almost overlapped, all parallel to the $x$ direction. In panels II and III, the fitted lines are no longer overlapped. Instead, while the angles included by the lines fitted by the PCA and the $x$ direction are 32.3° and 58.2°, which are close to 30° and 60°, the angles included by the lines fitted by the LS are 20.6° and 30.7°, respectively, notably deviated from the actually angles. Finally, in panel IV, while the angle included by the line fitted by the PCA and the $x$ direction is still close to 90°, the angle included by the line fitted by the LS is close to 0°, implying the failure of the LS method in such an extreme case.

Obviously, the directional lines fitted by the LS method could introduce considerable errors in the demodulation of the original signals.

Figs. 5(a) and 5(b) shows the waveforms and spectra of the 1.3-Hz signals demodulated with actual DC offsets from the noised sampling data plotted in Fig. 4 based on (6)~(8) of the DACM algorithm and (9)~(11) of the chord approximation, respectively. All the demodulated results were smoothed with the same moving average filter. The order of this moving average filter is 20. This low-pass filter with a cutoff frequency at 2 Hz was chosen because the frequency of the simulated sinusoidal heartbeat signal is only 1.3 Hz. Especially, the DACM algorithm was performed in two cases, with the actual DC offset located at (0, 0) and the ones optimized shown in Fig. 4, respectively.

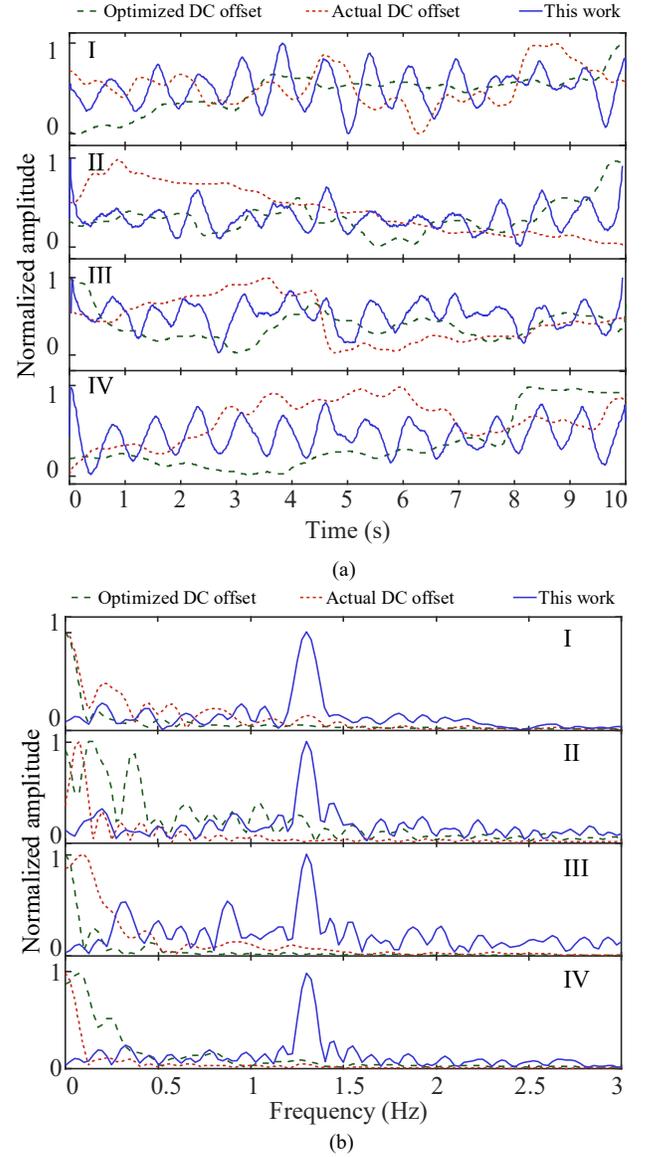

Fig. 5. Comparison between (a) the waveforms and (b) the spectra demodulated for the noised sampling data shown in Fig. 4 with different methods.

It is seen that even with the correct DC offsets, the signals demodulated by the DACM algorithm with both actual and optimized DC offsets are severely distorted. The 1.3-Hz signal cannot be observed in both time and frequency domains. It demonstrates that the demodulation algorithms based on the DC offset determination could become invalid in the present of strong noises.

In comparison, demodulated by the chord approximation, the 1.3-Hz signal can be observed in both the time and frequency domains with a satisfactory SNR. It verifies the theory discussed in Section II, showing a robust ability to demodulate signals from strongly noised data, even when the SNR of the raw quadrature data is only 3 dB.

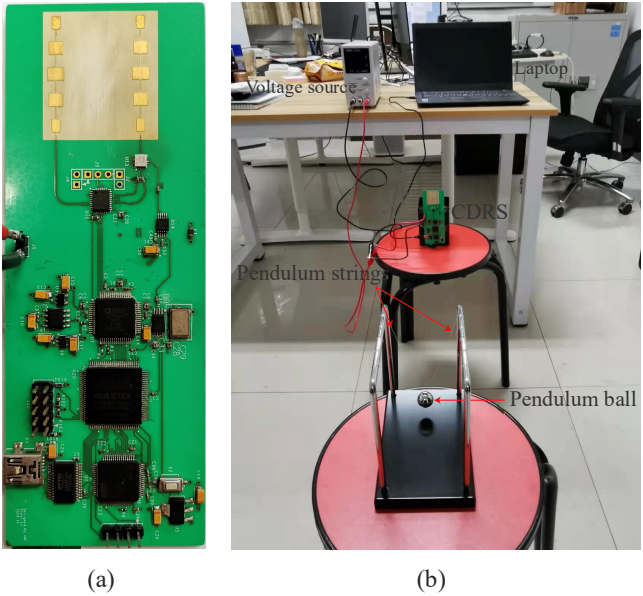

(a)                  (b)

Fig. 6. Experimental setup. (a) 24-GHz board-integrated CDRS used in the experiments. (b) Experimental setup for the pendulum experiment.

## IV. Experiment

In this section, experiments were performed to verify the practical performance of the proposed chord approximation. The experimental setup shown in Fig. 6 was established based on a board-integrated, 24-GHz CDRS developed in our previous work [20]. Based on a monolithic millimeter-wave radar chip, Infineon's BGT24MTR11 capable of transmitting 6-dBm power, and a pair of series-fed patch antennas with a 12-dBi gain, its performance for detecting human heartbeats has been verified.

### A. Pendulum experiment

To test the experimental setup and verify the proposed approach, the motion of a single pendulum was firstly detected taking advantage of its sinusoidal motion with a constant frequency and its easily controlled amplitude.

According to the pendulum equation, the period $T$ of its sinusoidal motion is determined by the length of the cycloid $h$, i.e.,

$$T = 2\pi\sqrt{l\cos\theta / g} = 2\pi\sqrt{h/g} \qquad (13)$$

where the gravitational acceleration g = 9.8 $N/kg$. In the experiment, $h$ was chosen as 14 cm, corresponding to a pendulum frequency of 1.33 Hz.

The experiment consists of two steps. In the first step, the CDRS was placed 40 cm away from the pendulum, and the field scattered by the swing pendulum was sampled by the CDRS. The experiment began when the amplitude of the pendulum swing was reduced to around 0.1 mm.

In the second step, after the sampling at the 40-cm distance, the CDRS was immediately moved backward to a place 60-cm away from the pendulum ball to sample the scattered field. The quadrature constellations consisting of sampled data are shown in Figs. 7(a) and 7(b), respectively. It is seen that due to the small swing amplitude and the small radar cross-section (RCS)

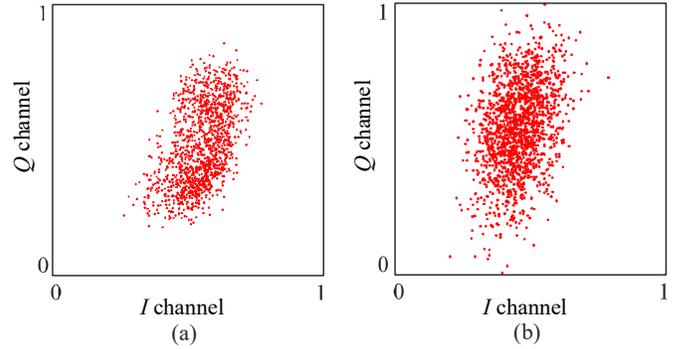

Fig. 7. Normalized raw data sampled in the pendulum experiment at distances of (a) 40 cm and (b) 60 cm, respectively. Both constellations are normalized by the amplitude of the data sampled at the distance of 60 cm,

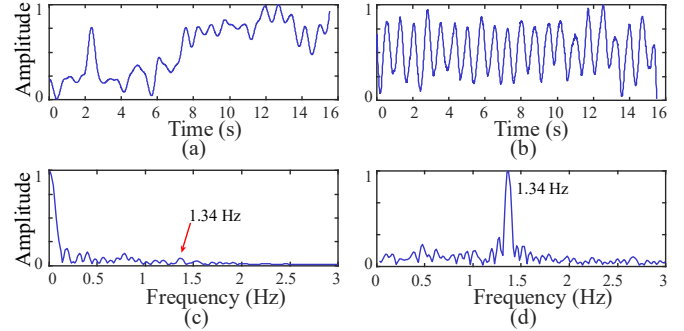

Fig. 8. Demodulation of the pendulum swings detected at the 40-cm distance. Subfigures (a), (b) and (c), (d) show the normalized waveforms and spectra demodulated by the DACM algorithm based on a DC-offset optimization and by the chord approximation, respectively.

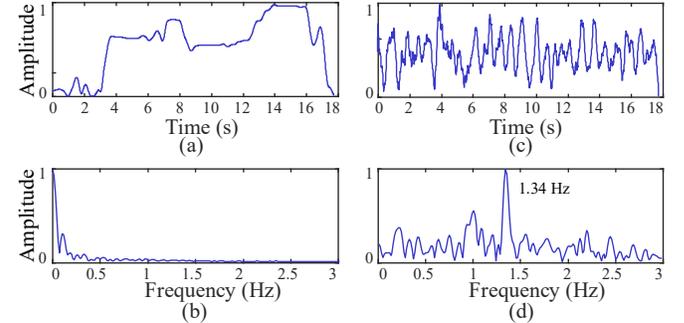

Fig. 9. Demodulation of the pendulum swing detected at the 60-cm distance.

of the 22-mm-diameter pendulum ball, the constellations look very noisy.

Fig. 8 shows the normalized waveforms and spectra of the pendulum swing demodulated by the DACM algorithm based on a DC-offset optimization and by the chord approximation from the raw data detected at 40-cm distance. It can be seen that demodulated by the DC offset optimization-based method, while the recovered waveform is severely distorted, the pendulum-swing spectrum appeared at 1.34 Hz is very small. In comparison, directly demodulated by the chord approximation algorithm, the quality of both the waveform and the spectrum are significantly improved.

Similarly, Fig. 9 shows the normalized waveforms and spectra of the pendulum swing demodulated at 60-cm distance, respectively. It can be seen that for the weakened signals, the optimization-based method failed to recover the pendulum swing spectrum. However, with the proposed chord approximation, the pendulum swing can still be demodulated, with a clear swing spectrum at 1.34 Hz.

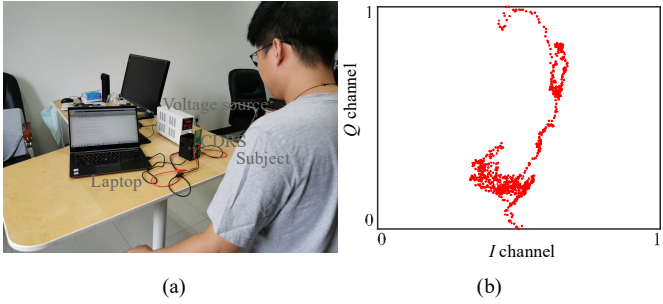

Fig. 10. Experiments for heartbeat detections. (a) Experimental environment. (b) Quadrature constellation of the detected heartbeat signal.

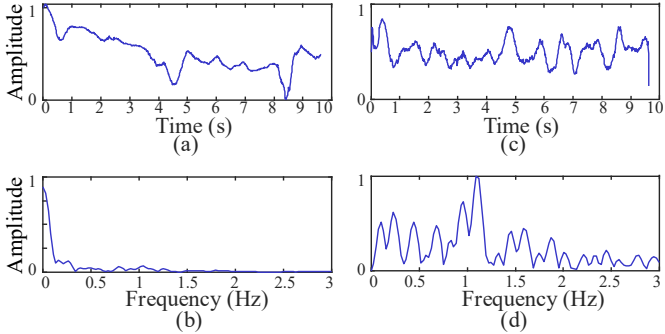

Fig. 11. Demodulation of the heartbeats. Subfigures (a), (b) and (c), (d) show the normalized waveforms and spectra demodulated with the DC-offset optimization and the chord approximation, respectively.

### B. Experimental heartbeat detection

Fig. 10(a) shows the experimental setup for heartbeat detections. In the experiment, a 32-year-old male sat in front of the CDRS with a distance around 50 cm and held his breath, to avoid the interference from his respirations. The detected quadrature constellation is shown in Fig. 10(b). Since there exist noises, the constellation is distorted, and thus does not look like an arc.

Fig. 11 shows the normalized waveforms and spectra of the heartbeats demodulated with the assistance of the DC-offset optimization and the chord-approximation without any consideration of DC offsets, respectively. Again, while the proposed method can robustly demodulate the waveform and the spectrum of the 1.22-Hz heartbeats, the DACM method failed to recover any heartbeat signal. Finally, calculated with a regular personal computer, the computation with and without the DC offset optimization costed 2.108 and 0.141 seconds, respectively.

To evaluate the absolute accuracy of the detected heartbeat rates, the heartbeats of three additional subjects identified as 1, 2 and 3 were detected at a 1-meter distance. Especially, their heartbeats were synchronously detected by a contact pulse sensor. Panels I, II, and III in Fig. 12(a) show the detected constellations, which were normalized by the one detected for subject 3. During the measurement, this subject slightly moved his body, resulting in a larger area of constellation and a moving

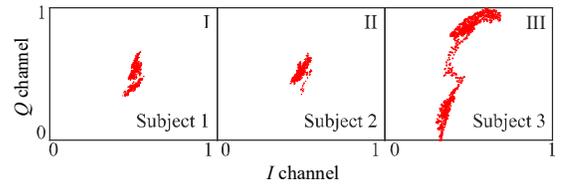

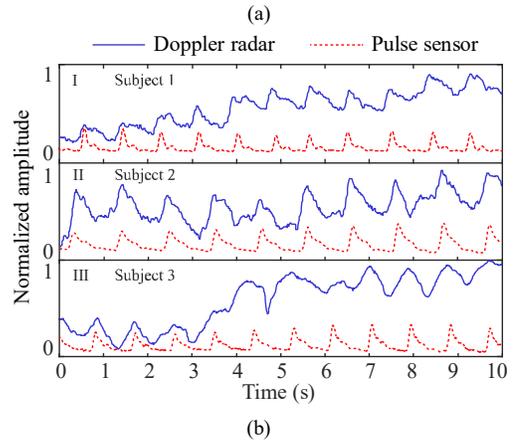

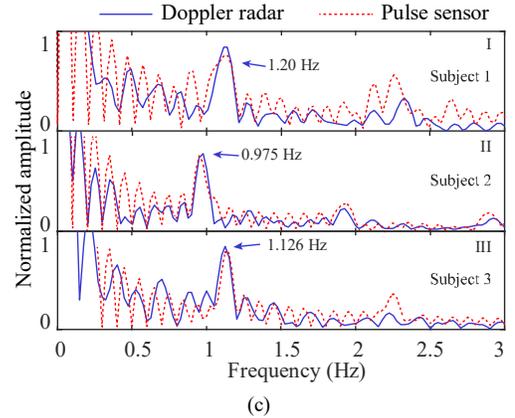

Fig. 12. Comparison between heartbeats simultaneously demodulated by the chord approximation and detected by the pulse sensor.

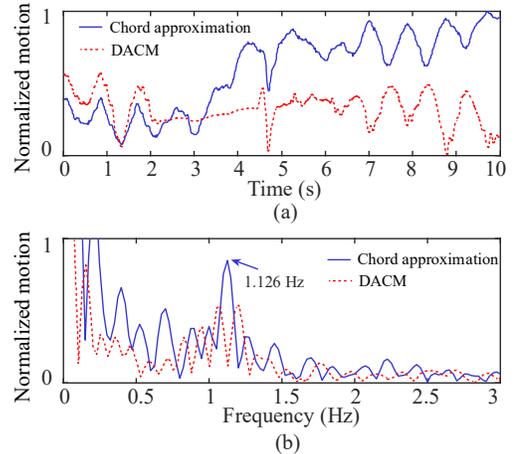

Fig. 13. Comparison between the demodulated results obtained by the DACM and the chord approximation for subject 3. (a) Time domain. (b) Frequency domain.

circle center.

Figs. 12(b) and 12(c) show the time- and frequency-domain heartbeats simultaneously demodulated by the chord approximation and detected by the pulse sensor. It is seen that for all the subjects, the demodulated heartbeats comply well with the ones detected by the contact pulse sensor. The average real-mean-square (RMS) error with respect to the heartbeat rate detected by the contact sensor is around 0.0152 Hz.

TABLE I
COMPARISON BETWEEN THE HEARTBEATS DETECTED BY THE
DACM, CHORD APPROXIMATION AND PULSE SENSOR

| Heartbeat rate (Hz) | Chord Approximation | DACM | Pulse sensor |
|---|---|---|---|
| Subject 1 (1 m) | 0.975 | 0.975 | 0.952 |
| Subject 2 (1 m) | 1.20 | 1.20 | 1.212 |
| Subject 3 (1 m) | 1.126 | Failure | 1.130 |

In addition, Fig. 13 shows the time- and frequency-domain heartbeats demodulated by the DACM and the chord approximation for subject 3. It is seen that the DACM fails to demodulate the heartbeat waveform during the movement of the circle center. As a result, the primary spectrum of the detected heartbeats is split, unable to determine the actual heartbeat frequency. In comparison, without needing to tracing the moving circle center, both the time- and frequency-domain heartbeats demodulated by the chord approximation are correct. All the related results are shown in Table I.

Fig. 14 shows the heartbeat detection in the presence of normal respirations. In practice, the displacement of the chest skin caused by the respiration is much larger than that caused by the heartbeat. The spectrum of the demodulated respiration signal will extend to the frequency region of the heartbeat, resulting in a deteriorated SNR of the heartbeat signal. Fig. 14 shows that although the interference due to the respiration is large, the heartbeat spectrum can still be identified, verifying the linearity of the chord approximation.

Finally, this experiment was repeated for subject 1 at distances increasing from 1 m to 5 m. It turned out that when the distance was increased to 2 m, the SNR of the detected amplitude of the heartbeat spectrum would reduce to 3 dB compared with its adjacent noise spectra. In practice, increasing the transmitting power and the gain of the antennas can help to increase this operating distance.

*C. Actuator experiment*

To quantitatively evaluate the accuracy of the chord approximation, a measurement to the sinusoidal motion of a copper plate precisely driven by an actuator was conducted, as shown in Fig. 15(a). The actual amplitude and frequency of this motion is 0.8-Hz and 0.2 mm, respectively, and the normalized amplitudes and frequency detected by the CDRS chord approximation are shown in Figs. 15(b) and 15(c), respectively.

It is seen that due to the very small motion scale, the normalized amplitude of the detected motion is perturbed. Compared with the actual amplitude and frequency, the real-mean-square (RMS) error of the detected frequency and the amplitude are 56 $\mu$m and 0.003 Hz, respectively, verifying the absolute accuracy of the chord approximation. This RMS error is estimated by the correspondence between the normalized and actual amplitudes plotted in Fig. 15.

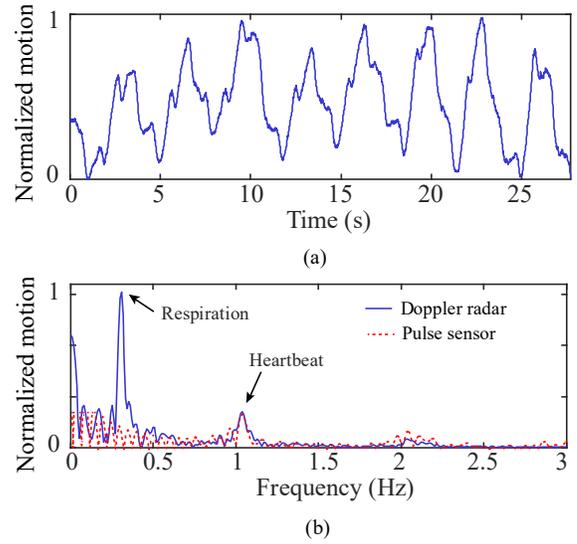

Fig. 14. Heartbeat detection in the presence of normal respirations.

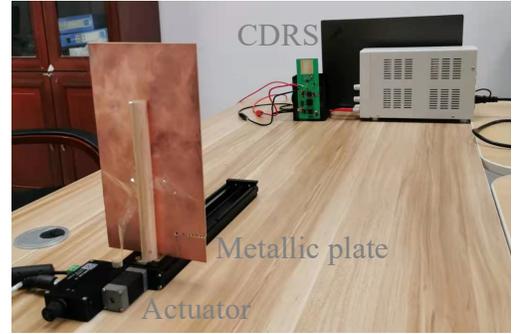

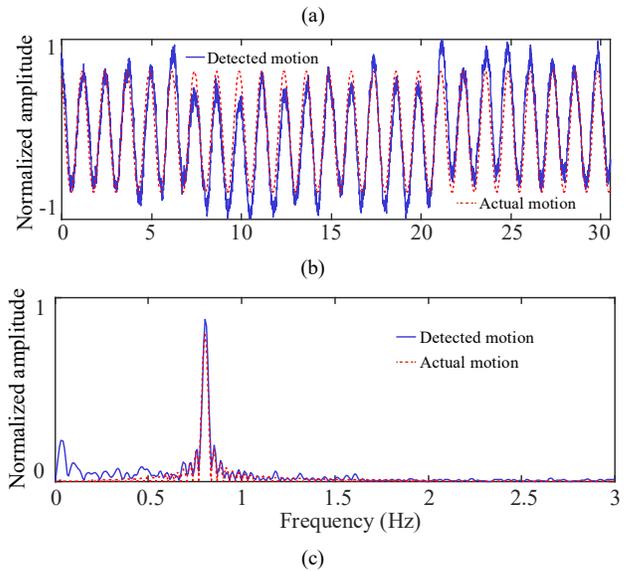

Fig. 15. Comparison between motions controlled by an actuator and detected by the chord approximation. (a) Experimental setup. (b) Normalized detected motion. (c) Spectra of the detected motions. In the normalization, the amplitude of the detected motion was normalized into -1 to +1. Then, the mean peak-to-peak amplitude of the detected motion was calculated, which is 1.35. Finally, the peak-to-peak amplitude of the actual sinusoidal motion was scaled to 1.35. Thus, the peak-to-peak amplitude of 1.35 corresponds to the actual amplitude of 0.2 mm.

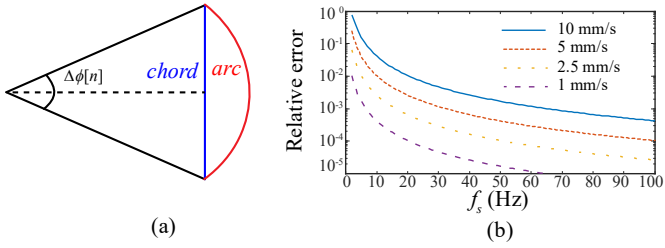

Fig. 16. Error analysis. (a) Analysis model. (b) Relative error caused by the chord approximation.

## D. Discussion

The above simulation and experimental results verified the effectiveness and accuracy of the proposed approach. They showed that while the anti-noise performance is significantly enhanced due to the chord approximation, the computation time used to demodulate heartbeat signals is reduced by an order of magnitude. Since it is no longer needed to optimize the circle centers to obtained the waveforms shown in Fig. 5, the average computation time cost by the DACM based on optimizing the circle center and by the proposed chord approximation are 1.188 and 0.315 seconds, respectively.

### 1) System error analysis

Mathematically, the chord approximation will result in a system error for the CDRSs. To analyze this theoretical error, we still denote the phase difference corresponding to an approximated chord as $\Delta\phi[n]$, as shown in Fig. 16(a). Then, the length of the arc and the chord can be calculated by

$$L_{arc} = R \cdot \Delta\phi[n] \quad (14)$$

$$L_{chord} = 2 \times \sin(\Delta\phi[n]/2) \times R \quad (15)$$

respectively. Then, the relative error can be defined as

$$error \triangleq (L_{arc} - L_{chord})/L_{arc} = 1 - 2\sin(\Delta\phi[n]/2)/\Delta\phi[n] \quad (16)$$

where

$$\Delta\phi[n] = 4\pi\Delta x(t)/\lambda = 4\pi v \Delta T/\lambda = 4\pi v/\lambda f_s \quad (17)$$

In (17), $\Delta x(t)$ is the motion between two adjacent sampling points, $\lambda$ is the wavelength of the carrier, $v$ represents the velocity of the moving object, $\Delta T$ is the sampling interval, and $f_s$ is the sampling rate.

Then, the relationship between the relative error, the sampling rate and the velocity of the object can be derived as

$$error = 1 - \sin(2\pi v/\lambda f_s)/(2\pi v/\lambda f_s) \quad (18)$$

It means that the system error due to the chord approximation depends on velocity $v$, wavelength $\lambda$ and sampling rate $f_s$.

Fig. 16(b) shows the relative error calculated based on (18) for velocities of different biomotions, in which $\lambda = 12.5$ mm, corresponding to the 24-GHz frequency. It is seen that for biomotion velocities of 1, 2.5, 5 and 10 mm/s, the relative error solely caused by the chord approximation drastically decreases with the increased sampling rate. As an example, with a 70-Hz sampling rate, the maximum error can be reduced to 0.1%. It means that for biomotions such as respirations and heartbeats, the errors caused by the chord approximation are negligible.

### 2) Determination of the points N in the segmented arcs

For a uniform motion with an amplitude of $A$, its Doppler phase can be expressed as

$$\phi = 4\pi A/\lambda = 4\pi vT/\lambda = 4\pi v N/f_s \lambda \quad (19)$$

where $\phi$ is the phase corresponding to a chord including $N$ sampling points, $\lambda$, $v$, $T$ and $f_s$ are the wavelength, velocity, moving time and sampling rate, respectively.

To satisfy the chord approximation condition, this arc should be shorter than a semicircle. In the meantime, to maintain an arc-like shape, its length can be larger than 15°. This condition can be expressed as

$$\pi/12 < \phi < \pi \quad (20)$$

Substituting (19) into (20), we obtain

$$\frac{\pi}{12} < \frac{4\pi v \cdot N/f_s}{\lambda} < \pi$$

and

$$\frac{\lambda \cdot f_s}{48v} < N < \frac{\lambda \cdot f_s}{4v} \quad (21)$$

It is seen that $N$ can be determined in a range by $\lambda$, $f_s$ and $v$ based on (21).

### 3) Limitation of the chord approximation

As indicated in the derivation of the chord approximation, the motion demodulated by this method is normalized. Therefore, the major limitation of the chord approximation is that the displacement, or absolute amplitude, of the biomotions cannot be given. It applies for biomotions whose temporal "patterns" are mainly concerned. It mainly applies for small-scale biomotions whose physical amplitudes are not concerned. Typical examples of such biomotions include human heartbeats and respirations.

For large-scale motions such as human walks, the same as the DACM method, the chord approximation still works by segmenting a large-scale motion into connected small-scale ones. Although only normalized motions can be detected, the obtained "patterns" of the large-scale motions can be used in applications such as the recognition and classification of human activities. It deserves in-depth investigations in the future.

Also, it should be noted that the PCA used in this work is different from that used in [27] for an FMCW radar. In [27], it works like a filter applied to the time-domain raw data, which could result in information loss of the detected signals. In comparison, the PCA used in this work is to fit the directional line. It does not intervene the signal processing, and thus will not result in any information loss of the raw data.

Finally, different from FMCW radars, the existing CDRS can only be used to detected a single biomotion. To solve this problem, the single-in multiple-out (SIMO) structured CDRS can be used to detected multiple biomotions. An example can be found in [28].

## V. CONCLUSION

In conclusion, we theoretically analyzed and experimentally verified that a DC-offset-immune demodulation algorithm that can be used to replace the conventional methods based on optimization of DC offsets that are unavoidable in CDRSs. It approximates the arcs in the quadrature constellations as chords, very suitable for detecting weak biomotions such as human heartbeats. Assisted by this novel method, while the time-consuming DC offset determination and tracing can be completely avoided, the anti-noise performance of CDRSs can be significantly improved. By segmenting the constellation signals, the proposed chord approximation can also be used to detect large-scale biomotions. It deserves in-depth investigations in the future. We envision wide applications of this robust method in agile motion detections such as the biomotion detections.